\documentclass{article}

\usepackage{PRIMEarxiv}

\usepackage[utf8]{inputenc} 
\usepackage[T1]{fontenc}    
\usepackage{hyperref}       
\usepackage{url}            
\usepackage{booktabs}       
\usepackage{amsfonts}       
\usepackage{nicefrac}       
\usepackage{microtype}      
\usepackage{lipsum}
\usepackage{fancyhdr}       
\usepackage{graphicx}       
\usepackage{xcolor}
\usepackage{comment}
\usepackage{subcaption}

\title{Uncovering IP Address Hosting Types Behind Malicious Websites
}

\author{
  Nimesha Wickramasinghe$^*$, Mohamed Nabeel$^+$, \\Kenneth Thilakaratne$^*$, Chamath Keppitiyagama$^*$ , Kasun De Zoysa$^*$ \\
  $^*$University of Colombo School of Computing \\
  $^+$Qatar Computing Research Institute
}

\begin{document}
\maketitle

\begin{abstract}
Hundreds of thousands of malicious domains are created by miscreants everyday. These malicious domains are hosted on a wide variety of network infrastructures. Traditionally, attackers utilize bullet proof hosting services (e.g. MaxiDed, Cyber Bunker) to take advantage of relatively lenient policies on what content they can host. However, these IP ranges are increasingly being blocked or the services are taken down by law enforcement. Hence, attackers are moving towards utilizing IPs from regular hosting providers while staying under the radar of these hosting providers. There are several practical advantages of accurately knowing the type of IP used to host malicious domains. If the IP is a dedicated IP (i.e. it is leased to a single entity), one may blacklist the IP to block domains hosted on those IPs as welll as use as a way to identify other malicious domains hosted the same IP. If the IP is a shared hosting IP, hosting providers may take measures to clean up such domains and maintain a high reputation for their users.
\end{abstract}


\section{Introduction}\label{sec:intro}
Malicious websites rely on hosting infrastructures to launch their attackers. 
Traditionally, attackers utilized Bulletproof Hosting Services (BPHS) such as MaxiDed and Cyber Bunker to take advantage of their relatively lenient policies on what content they can host \cite{Goncharov:2015:BPHS}. However, these IP address (IP) ranges are increasingly being blocked or the services are taken down by law enforcement \cite{ko:Security:2019}. Hence, attackers are now moving towards utilizing IPs from regular hosting \\providers while staying under the radar of these service providers \cite{Pa:2015:nameServerBased}. Further, public hosting provides economies of scale as well as allows to launch attacks at scale from multiple geographic locations. The lack of considerable effort to collect information about the users who register with the service providers nor having proper security mechanisms to detect or prevent these attacks make the life of attackers easier~\cite{HostingProviders}. It is, however, not clear to what extent this transition from BPHS to regular hosting services has been occurring. 

Blocklists are a widely used network security mechanism to block undesired network traffic based on IP reputation~\cite{blag2020}. Blocklists can be applied at various points in a security architecture, such as a host, web proxy, DNS servers, email server, firewall, directory servers or application authentication gateways \cite{strategicBased}. While they are useful, such blocklists may result in high collateral damage if the blocked IPs are co-hosting benign domains. Current blocklists do not make a distinction between IP addresses solely used for malicious activities and IP addresses that co-host benign domains.


\begin{figure*}[!th]
\centering
    \includegraphics[width=\textwidth]{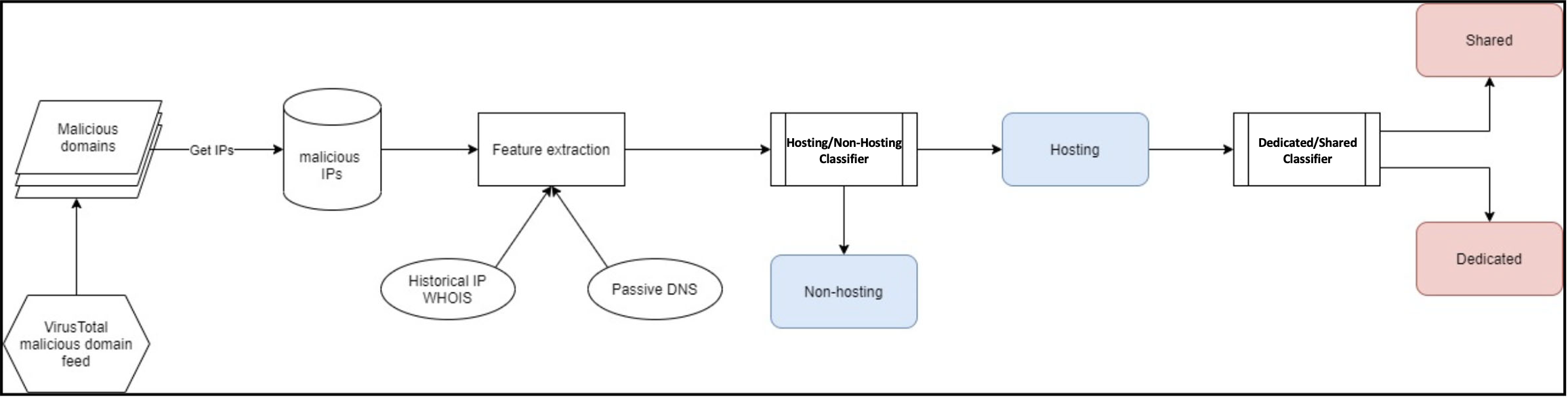}
    \caption{Overall Pipeline of Identifying Hosting Types of Malicious Domains}
    \label{fig:overall}
\end{figure*}

Domain threat intelligence services rely on hosting IPs of malicious domains to identify additional malicious domains controlled by attackers~\cite{inferenceGraph}. Such co-hosting relationships yield favorable results only if those IP addresses host domains belonging to the same entity. Identifying if an IP is associated with domains belonging to one entity is a challenging task due to several reasons: (1) one needs to identify \emph{all} the domains hosted on the IP address, (2) ownership of a domain is often times not available or difficult to find, and (3) the domains hosted IPs change over time.

While the hosting type of an IP address is quite useful in practice, it is not readily available except for selected cases. For example, major cloud service providers such as Amazon AWS~\cite{AWSPublicIP}, Microsoft Azure~\cite{AzurePublicIPList} and Google GCE~\cite{GooglePublicIPAPI}, have made their IP address ranges public, but there is a long tail of other hosting providers whose IP ranges are not readily available. Further, third-party services that provide such information are either severely throttled or quite expensive~\cite{ipinfo}. 

In this work, we design machine learning models to solve the problem of identifying the hosting type of an IP address. We first make the distinction between \emph{hosting} and \emph{non-hosting} IP addresses. A hosting IP address belongs to a hosting service and anyone may utilize, often with a usage fee, to host their domains whereas a non-hosting IP, as the name suggests,  are either not leased to third-party users or residential IP addresses belonging to ISPs. We further categorize hosting IP addresses as either \emph{shared} or \emph{dedicated}. A shared hosting IP (shared IP for short) hosts domains belonging to more than one user or organization, whereas a dedicated hosting IP (dedicated IPs for short) hosts domains belonging to a single user or organization.

Based on the trained classifiers, we study the properties of IP addresses hosting malicious domains collected from VirusTotal URL feed~\cite{VirusTotal}. Our study identifies that a vast majority (95.2\%) of malicious websites are hosted on regular hosting IPs and 97.1\% of these malicious websites are co-hosted with other unrelated benign websites.

We make the following contributions in this work:
\begin{itemize}
    \item We build a machine learning model to identify hosting IPs from non-hosting IPs with a precision of 98.03\%, recall of 96.89\% and false positive rate (FPR) of 1.95\%.
    \item We build another machine learning model to identify shared IP from dedicated IPs in the hosting IP category with a precision of 94.22\%, recall of 94.00\% and FPR of 5.77\%.
    \item We study the hosting IP types of IPs hosting malicious domains and show that overwhelming majority of malicious domains utilize shared hosting environments.
\end{itemize}

\section{Definitions}\label{sec:background}
In this section, we formally define the different hosting types. 
Recall that we first categorize all IPs in the wild into hosting and non-hosting IPs. Hosting IPs are further categorized into shared and dedicated IPs.

\textbf{Hosting IP}: An IP address that belongs to a hosting provider which allows third parties to host on (E.g.: Amazon AWS~\cite{aws})

\textbf{Non-hosting IP}: An IP address that is not publicly offered to third-parties to host. These IPs typically include IP ranges of ISPs who usually lease them to businesses or use them as resdiential IP blocks (E.g.: Verizon residential IP addresses), and of large organizations/universities (E.g.: Facebook, MIT). 

\textbf{Dedicated IP}: - A hosting IP that hosts domains belonging to one single entity. An entity can be either a user or an organization. 

\textbf{Shared IP}: A hosting IP that hosts domains belonging to two or more entities. 

\section{Study Design}\label{sec:design}
In this section, we outline our study methodology and data collection.

\vspace{-3mm}
\subsection{Methodology}
We first collect hosting and non-hosting IPs distributed across the world. Then, our domain experts label a subset of hosting IPs as shared and dedicated. Hosting classifier takes a ground truth set of IPs along with their features, and builds a machine learning model to classify IPs as hosting and non-hosting. Shared/Dedicated classifier takes a ground truth set of hosting IPs to build a machine learning model to classify IPs as shared and dedicated. Figure~\ref{fig:overall} shows the final pipeline of identifying the hosting types of malicious domains based on the above mentioned two trained classifiers. 

\vspace{-3mm}
\subsection{Data Collection and Labeling}

In order to study and understand discriminative features, 400,000 IP addresses are collected, including 200,000 non-hosting IPs and 200,000 hosting IPs. The next section comprises the steps we follow to collect these IP addresses. 

\subsubsection{Non-hosting IP Collection}\hfill
We first collect a tentative diverse set of non-hosting IPs and perform additional checks to ascertain that they are in fact non-hosting.

Universities, government and military organizations are highly likely to own huge subsets of IP addresses which are not used for web hosting~\cite{inferenceGraph}. We extract IP blocks owned by these institution by keyword matching organization from IP-ASN database from the MaxMind DB. We select a random subset of IPs from this collection and gather domains hosted on these IP address utilizing Farsight Passive DNS service. We extract those those IPs whose owner is matching with the owner of the websites. In order to diversify the non-hosting IPs, we also include 20,000 Residential IPs obtained from a prior work~\cite{revil:2019}. Overall, we collect 200,000 non-hosting IP addresses.

\vspace{-3mm}
\subsubsection{Hosting IP Collection}
We collect hosting IP addresses from publicly advertised hosting IP ranges from major hosting provides such as GCP, AWS and Azure, and from MaxMind IP-ASN database for web hosting providers such as Dream Host, Host Monster, GoDaddy, Bluehost, Host Dime, Host Gator, Liquid web and Host Europe. 
From nearly 80 million publicly advertised hosting IP addresses, we selected a random sample of 200,000 to conduct our study.

\vspace{-3mm}
\subsubsection{Dedicated and Shared IP Collection}
From the collected hosting IPs, we identify a ground truth set of dedicated and shared IPs. We first collect the domains hosted on each of these IP addresses using the Passive DNS service. We then collect WHOIS records for these domains. It should be noted, as observed in prior research~\cite{whois:imc:2016}, we are only able to collect WHOIS record for a limited number of domains and the registrant in most of these collected WHOIS records are either redacted or privacy protected.
This limits our ability to determine the domain ownership solely using WHOIS records. Hence, we follow the following procedure to identify dedicated and shared IP addresses.


\begin{itemize}
    \item If an IP is hosting only one domain, it is by definition a dedicated IP.
    \item If WHOIS records are available and the registrants are not redacted or privacy protected, we mark those hosting IPs having same registrant for all domains as dedicated and otherwise shared.
    \item If all websites hosted on a given IP redirects to the same parent website, we mark it as dedicated as they are all control by the same entity.
    \item For the remaining, we visually inspect the web page content to find similarity in terms of information included, trademark and contact information. If all websites hosted on an IP has similar content, we mark the IP as dedicated.
\end{itemize}

\subsubsection{Data Sources}
We utilize Farsight Passive DNS service~\cite{DNSDB} to extract historical hosting information about domains and IPs. For the 400,000 IP addresses, we collected nearly 45 million Passive DNS records in Aug. 2020 and in Dec. 2020. In order to diversify our findings, we collected Passive DNS records across different time intervals. VirusTotal URL Feed~\cite{VirusTotal} on the other hand, is used to identify malicious domains for this study. VirusTotal provides an aggregated report for each URL containing results from 70+ scanners.  Following best practices~\cite{compromised2021}, we select those domains with at least 5 scanners detecting positive as our malicious domain corpus. Maxmind database~\cite{maxmind} is utilized to extract IP ASN information.

\section{Classifying IP Addresses}\label{sec:classify}
\subsection{Feature Engineering}\label{subsec:featureEngineering}
According to our classification hierarchy, the first step is to identify features with high discriminative power to identify if a given IP is a Hosting IP or a Non-hosting IP. The second step is to identify features to classify the hosting IPs as dedicated or shared. The identified features for the two steps are describe in this section.
\vspace{-3mm}
\subsubsection{Features for the Hosting/Non-hosting IP Classifier}\label{subsec:features}

\begin{table*}[!th]
\caption{Hosting/Non-hosting classification features}
\label{tab:hosting}
\begin{minipage}{\linewidth}
\centering
\footnotesize
\begin{tabular}{| p{0.7cm} | p{2.7cm} | p{9.5cm} | p {1.7cm} |}
\hline
\multicolumn{1}{|c|}{\textbf{ID}} &
\multicolumn{1}{|c|}{\textbf{Feature Name}} &  \multicolumn{1}{c|}{\textbf{Description}} &
\multicolumn{1}{c|}{\textbf{Type}} 

\\ \hline
1
&
number of TLD+2
&
Number of TLD+2 domains resolved to the given IP 
&
Numerical
\\ \hline

2
&
number of TLD+3
&
Number of TLD+3 domains resolved to the given IP
&
Numerical
\\ \hline

3
&
number of domains
&
Number of domains resolved to the given IP
&
Numerical
\\ \hline

4
&
percent DNS
&
Percentage of IPs in /24 IP prefix with DNS records
& 
Numerical
\\ \hline

5
&
mean TLD+3
&
Mean number of TLD+3 domains resolved to IPs in /24 IP prefix
&
Numerical
\\ \hline

6
&
max TLD+3
&
Maximum number of TLD+3 domains resolved to IPs in /24 IP prefix
&
Numerical
\\ \hline

7
&
mean TLD+2
&
Mean number of TLD+2 domains resolved to IPs in /24 IP prefix
&
Numerical
\\ \hline

8
&
max TLD+2
&
Maximum number of TLD+2 domains resolved to IPs in /24 IP prefix 
&
Numerical
\\ \hline

9
&
number of owners 
&
Number of owners of the IP during the past 10 years
& 
Numerical
\\ \hline

10
&
number of inetnums
&
Number of unique historical inetnums (IP ranges) during the past 10 years
&
Numerical
\\ \hline

11
&
max inetnum size
&
Maximum size of the inetnums during the past 10 years
&
Numerical
\\ \hline

12
&
min inetnum size
&
Minimum size of the inetnums during the past 10 years
&
Numerical
\\ \hline

13
&
size of inetnum
&
Size of the inetnums during the past 10 years
&
Numerical
\\ \hline

14
&
type of inetnum
&
Assignment type of the current inetnum
&
Categorical
\\ \hline

15
&
most recent update
&
Most recent WHOIS record update
&
Numerical
\\ \hline

16
&
number of whois
&
Number of WHOIS records during the past 10 years
&
Numerical
\\ \hline

\end{tabular}
\end{minipage}
\end{table*}

Table~\ref{tab:hosting} shows the features we utilize to train the hosting/non-hosting classifier. We provide a brief intuition and observations of these features below. Number of TLD+2/TLD+3/all domains resolved to a given IP describe the number of second level, third level and all domains hosted on a given IP. Intuitively, as hosting providers need to host many domains in a limited range of IP addresses compared to Non-hosting IP owners, it is highly likely for Hosting IPs to host more domains per IP. The evaluation on the labeled set shows that Non-hosting IPs host 1.06, 0.56, 2.07 domains on average for the three categories whereas Hosting IPs host 452.6, 40.3, and 691.45 domains.

Features 4 to 8 extract statistics of domains resolved to IPs in /24 IP prefix. Since IPs are allocated to different ISPs in subnets, it is highly likely for /24 IP prefixes of Hosting IPs to have a higher number of domains mapped compared to the number of /24 IP prefixes of Non-hosting IPs.

Features 9 to 14 are related to inetnums. An inetnum object describes a range of IP addresses and all its attributes. It is known that hosting IPs are resold and reallocated more often to different vendors [9] compared to non-hosting IPs whose owners are fairly stable. This observation is utilized and the historical ownership of IPs (Feature 9) to differentiate between the two IP types. In the ground truth IPs, it is found that the average number of historical owners for Hosting IPs is 6.8 whereas it is 1.2 for Non-hosting IPs. All inetnums are created in a hierarchical manner and therefore it forms an inetnum tree. Since Feature 9 shows a significant difference, it is likely the allocated inetnums  differ too and this behavior is captured with Feature 10. Features 11 to 13 represent maximum size/minimum size/size of the inetnums respectively. Residential IPs under non-hosting IPs tend to have much more objects in inetnums. 
Though there is no major difference between the maximum and minimum number: the size difference is significant. For non-hosting IPs, the average size of inetnum is 15,893,510.4 and 8,652,824.7 for Hosting IPs. Feature 14 identifies the assignment type of the current direct inetnum. WHOIS record of an IP address explains the NET TYPE of a given IP address. The categories are: Direct Allocation (ISP), Direct Assignment (End user), Reallocated (downstream ISP customer), and Reassigned (end user customer). It is observed that hosting IPs are mostly in the Direct Allocation category as they are allocated for other hosting providers by Regional Internet Registries.

Features 15 and 16 are related to the timestamps of IP WHOIS records. As Hosting IPs are changing hands often, intuitively they are likely to have more recent last update dates. Feature 15 capture this attribute.  The average last update date for Non-hosting IPs is on average 7 years whereas the same for hosting IPs is 2 years. Feature 16 measures the number of WHOIS records of the last 10 years. The number of WHOIS records for Hosting IPs is likely to be more than that for Non-hosting IPs as hosting IPs change frequently. We observe from our data that the average number of WHOIS records associated with a Hosting IP is around 3.67 whereas it is 1.26 for a Non-hosting IP.

\subsubsection{Features for Dedicated/Shared IP Classifier}

\begin{table*}[!th]
\caption{Dedicated/Shared classification features}
\label{tab:dedicated}
\begin{minipage}{\linewidth}
\centering
\footnotesize
\begin{tabular}{| p{1cm} | p{3cm} | p{8cm} | p {2cm} |}
\hline
\multicolumn{1}{|c|}{\textbf{ID}} &
\multicolumn{1}{|c|}{\textbf{Feature Name}} &  \multicolumn{1}{c|}{\textbf{Description}} &
\multicolumn{1}{c|}{\textbf{Type}} 

\\ \hline
1
&
number of TLD+2
&
Number of TLD+2 domains resolved to the given IP 
&
Numerical
\\ \hline

2
&
number of TLD+3
&
Number of TLD+3 domains resolved to the given IP
&
Numerical
\\ \hline

3
&
number of domains
&
Number of domains resolved to the given IP
&
Numerical
\\ \hline

4
&
number of owners
&
Number of owners of the IP during the past 10 years
&
Numerical
\\ \hline

5
&
number of whois
&
Number of WHOIS records in the past 10 years
&
Numerical
\\ \hline

6
&
avg daily churn
&
The average daily churn of domains hosted on the IP
&
Numerical
\\ \hline

7
&
std daily churn
&
The standard deviation of daily churn
&
Numerical
\\ \hline

8
&
avg duration
&
The average duration of the domains hosted on the IP
&
Numerical
\\ \hline

9
&
std duration
&
The standard deviation of the duration of the domains
&
Numerical 
\\ \hline

\end{tabular}
\end{minipage}

\end{table*}

In this section, we describe the features we utilize for dedicated/shared IP classifier. We find that some of the features utilized with hosting/non-hosting classifier are quite useful for this classifier. Specifically, as shown in Table~\ref{tab:dedicated}, features 1 to 5 are reused but their proportions are different in the ground truth of dedicated and shared IPs.

Intuitively, Shared IPs need to host a higher number of domains belonging to different entities compared to Dedicated IPs. The evaluation on the labeled set shows that Dedicated IPs host, on average 796.08, 43.25, 539.828 second, third and all level domains whereas Shared IPs host, on average, 2624.18, 609.88, 1650.06 for the same. 

Features 6 and 7 represent Mean and Standard deviation of the daily churn of domains hosted on a given IP address. Shared IPs are likely to have more churn compared to the dedicated IPs due to its changing nature. We collect daily hosted apexes on each IP for the last 60 days and measure the mean and the standard deviation. The average daily churn for Dedicated IPs is 0.12, whereas it is 1.18 for Shared IPs. The average standard deviation is 0.68 and 5.91 for Dedicated and Shared IPs respectively.

Features 8 and 9 measures the Mean and Standard deviation of the duration of domains hosted on a given IP address. Domains hosted on Shared IPs are likely to be more ephemeral and likely to vary more across hosted domains compared to those on Dedicated IPs. Our analysis shows that the average duration of all the Dedicated IPs is 2.3 years and it is 1.2 years for Shared IPs. The standard deviation for these two classes are 1.4 and 1.1 respectively.

\vspace{-2mm}
\subsubsection{Ground Truth and Feature Extraction}\label{subsec:featureext}
We utilize Passive DNS and historical IP WHOIS data to extract the features described in Setion~\ref{subsec:features}. Features 1-8 in hosting/non-hosting classifier and Features 1-3 in dedicate/shared classifier are derived from passive DNS. The rest of the features are derived from historical IP WHOIS data.

In Jan. 2021, we collect features for two ground truth data sets H-GT1 and H-GT2 for Hosting/Non-Hosting IP classifier. IPs for these two data sets were selected randomly from the 400,000 IPs we studied initially. Each set has 1000 IPs from each class along with the identified features. Similarly, we collect features for two ground truth datasets D-GT1 and D-GT2 for Dedicated/Shared IP classifier. Each set has 400 IPs from each class.

\subsection{Training Machine Learning Models}~\label{model}
We train 8 classifiers (Support Vector Classification (SV), Random Forest (RF), Extra Tree (ET), Logistic Regression(LR), Decision  Tree  (DT), Gradient  Boosting  (GB), AdaBoosting (AB) and K-Neighbors (KN) Classification) using the features Section~\ref{subsec:featureEngineering}. We test all these classifiers using 5-fold cross validation techniques and realized that RF outperform others in our task.  
Hence, in the rest of the paper, we present results only from RF classifier. 

\begin{figure}[!ht]
\centering

\begin{subfigure}{0.45\linewidth}
\includegraphics[width=\linewidth]{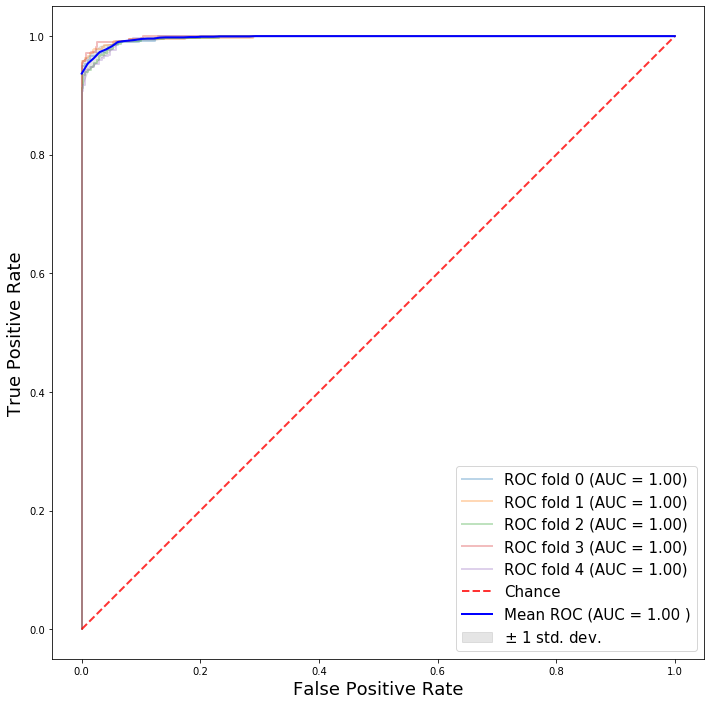} 
\caption{H-GT1}
\label{fig:hostingroc1}
\end{subfigure}
\hfill
\begin{subfigure}{0.45\linewidth}
\includegraphics[width=\linewidth]{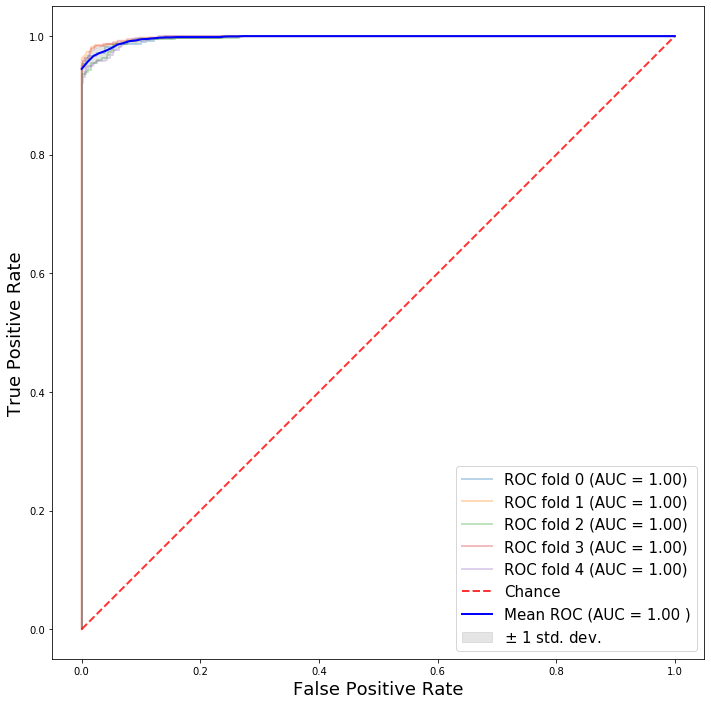} 
\caption{H-GT2}
\label{fig:hostingroc2}
\end{subfigure}

\caption{ROC Curves for Hosing/Non-hosting Classifier for the Two Ground Truth Sets} 
\label{fig:hostingroc}
\end{figure}

\begin{table}
\centering
\caption{Performance of Hosting/Non-Hosting Classifier for the Two GT Sets}
\label{tab:hostingperf}
\begin{tabular}{| p{0.41in} || p{0.72in} | p{0.735in} | p{0.735in} |}
\hline
\small{\textbf{Dataset}} & \small{\textbf{Precision}} & \small{\textbf{Recall}} & \small{\textbf{FPR}}\\
\hline
\hline
H-GT1 & 97.98 & 96.68 & 2.00 \\ 
 \hline
H-GT2 & 98.03 & 96.89 & 1.95 \\
 \hline
\end{tabular}
\end{table}

\begin{figure}[b]
\centering
    \includegraphics[scale=0.45]{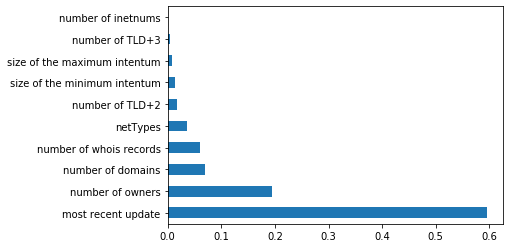}
    \caption{Feature Importance for Hosting/Non-Hosting Classifier}
    \label{fig:hostingimp}
\end{figure}

Figure~\ref{fig:hostingroc} shows the ROC curves for the two ground truth datasets for Hosting/Non-Hosting Classifier. We observe consistently high performance across the two datasets. Table~\ref{tab:hostingperf} shows the performance of the classifier. Our classifier achieves a precision of 98.03\%, a recall of 96.89\% and FPR (false positive rate) of 1.95\%. Figure~\ref{fig:hostingimp} shows the importance of each feature. The most recent WHOIS update, the number of owners and domains are the most discriminating features.

\begin{figure}[!ht]
\centering
\begin{subfigure}{0.45\linewidth}
\includegraphics[width=\linewidth]{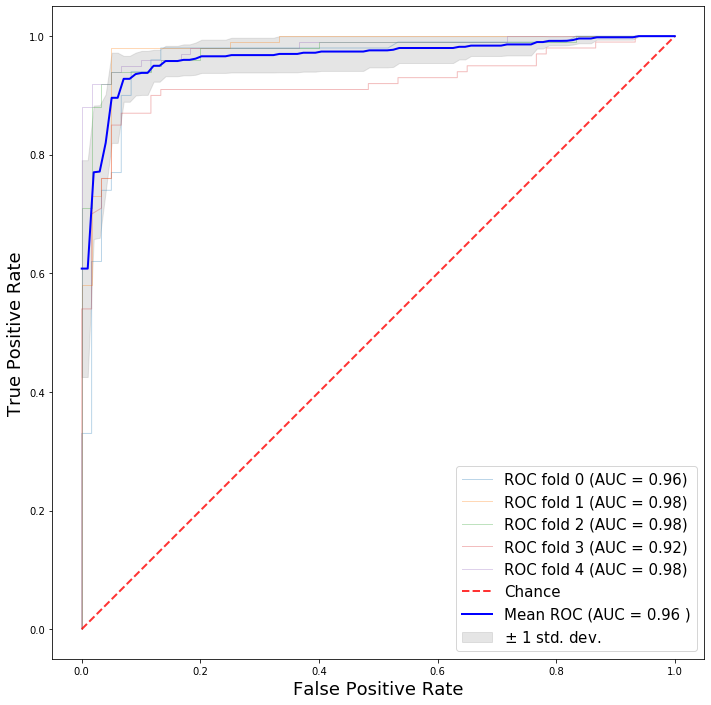} 
\caption{D-GT1}
\label{fig:dedicatedroc1}
\end{subfigure}
\hfill
\begin{subfigure}{0.45\linewidth}
\includegraphics[width=\linewidth]{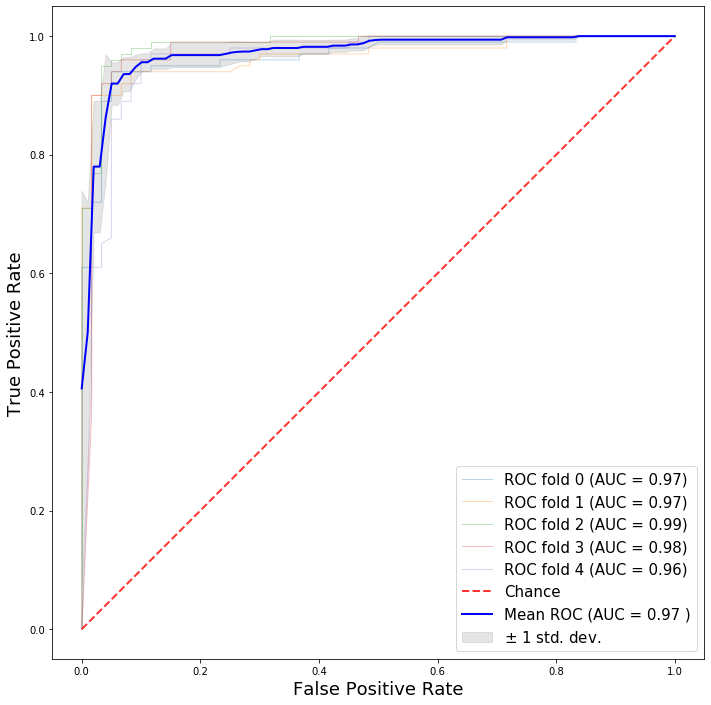} 
\caption{D-GT2}
\label{fig:dedicatedroc2}
\end{subfigure}
\caption{ROC Curves for Dedicated/Shared Classifier for the Two Ground Truth Sets} 
\label{fig:dedicatedroc}
\end{figure}

\begin{table}
\centering
\caption{Performance of Dedicated/Shared Classifier for the Two GT Sets}
\label{tab:dedicatedperf}
\begin{tabular}{| p{0.41in} || p{0.72in} | p{0.735in} | p{0.735in} |}
\hline
\small{\textbf{Dataset}} & \small{\textbf{Precision}} & \small{\textbf{Recall}} & \small{\textbf{FPR}}\\
\hline
\hline
D-GT1 & 95.00 & 94.4 & 8.33 \\ 
 \hline
D-GT2 & 94.22 & 94.00 & 5.77 \\
 \hline
\end{tabular}
\end{table}

\begin{figure}[!]
\centering
    \includegraphics[scale=0.45]{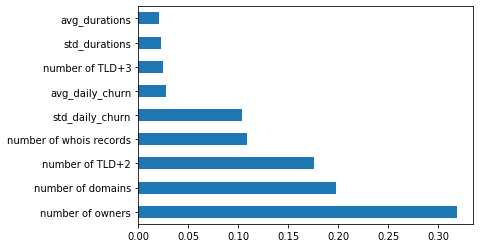}
    \caption{Feature Importance for Dedicated/Shared Classifier}
    \label{fig:dedictedimp}
\end{figure}

Figure~\ref{fig:dedicatedroc} shows the ROC curves for the two ground truth datasets for Dedicated/Shared Classifier. We observe consistently high performance across the two datasets. Table~\ref{tab:dedicatedperf} shows the performance of the classifier. Our classifier achieves a precision of 94.22\%, a recall of 94.00\% and FPR (false positive rate) of 5.77\%. Figure~\ref{fig:dedictedimp} shows the importance of each feature. The number of owners, domains and TLD+2s are the most discriminating features.

\begin{table*}[!]
\caption{Top Hosting Providers with the Highest Number of Benign and Malicious Domains}
\label{tab:hosting}
\begin{minipage}{\linewidth}
\centering
\footnotesize
\begin{tabular}{| p{3cm} | p{2cm} || p{3cm} | p {1cm} || p{3cm} | p {1cm} |}
\hline
\multicolumn{1}{|c|}{\textbf{All Hosting}} &
\multicolumn{1}{|c||}{\textbf{\#AllDomains}} &  
\multicolumn{1}{|c|}{\textbf{Shared Hosting}} &
\multicolumn{1}{|c||}{\textbf{\#MalDomains}} &
\multicolumn{1}{|c|}{\textbf{Dedicated Hosting}} &
\multicolumn{1}{|c|}{\textbf{\#MalDomains}}
\\ \hline

Wix.com Ltd & 18,411,677 & Cloudflare Net & 22,473 & Amazon-02 & 2,126
\\ \hline

Google & 14,679,240 & Amazon-02 & 14,139 & Cloudfare Net & 194
\\ \hline

Cloudflare Net & 12,786,921 & Namecheap & 654 & DigitalOcean & 123
\\ \hline

GoDaddy Inc. & 11,114,352 & Unified Layer & 450 & OVH & 68
\\ \hline

Amazon-02 & 9,948,052 & Hostinger & 249 & Hostinger & 65
\\ \hline
\end{tabular}
\end{minipage}
\end{table*}

\section{Hosting Types of Malicious Domains}\label{sec:analysis}
Using the trained classifiers, we analyze the hosting types of 174,701 malicious domains collected from VirusTotal URL Feed in January 2021. We first identify the hosting IP addresses for these websites. These domains are hosted on 73,441 IP addresses. We are able to collect features for 47,141 IP addresses. Hence, we study the hosting types of these latter set of IPs where 143,316 malicious domains are hosted.

We first use the trained Hosting/Non-hosting classifier to predict hosting and non-hosting IPs in this set. With 95\% confidence, we identify 44,890 (95.2\%) hosting IPs and 2,251 (4.8\%) non-hosting IPs. Out of the hosting IPs, our Dedicate/Shared classifier detects 43,605 (97.1\%) shared IPs and 1,285 (2.9\%) dedicated IPs.

\begin{figure}[!]
\centering
    \includegraphics[scale=0.45]{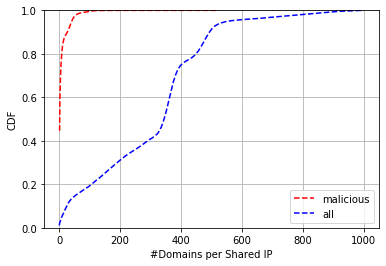}
    \caption{Number of Malicious and Overall Domains per Shared IP}
    \label{fig:sharedhosting}
\vspace{-5mm}
\end{figure}

Figure~\ref{fig:sharedhosting} shows the distribution of malicious and all domains on shared IPs. It shows that while there are more than 200 domains hosted on 70\% of shared IPs, nearly 50\% of these IPs host a handful of malicious domains. This shows that attackers utilize large pool of unrelated domains making it difficult to identify and block/take down. They also tend to choose shared hosting due to to their low cost of renting and their poor security measures \cite{zeroDay}.  
Hence, in the rest of the paper, we present results only from RF classifier. 

Table~\ref{tab:hosting} shows the top 5 hosting providers in terms of total number of domains hosted, total number of malicious domains hosted on shared IPs and the total number of malicious domains hosted on dedicated IPs. It is concerning to observe that Cloudflare and Amazon have the highest number of malicious domains hosted under both shared and dedicated settings. Google as a hosting service provider, on the other hand, has much less number of malicious domains in their infrastructure. We believe this information is quite useful in practice to make business decision to utilize either dedicated or shared hosting and under which hosting provider.

\section{Related Work}\label{sec:related}
One may wonder if published IP address ranges could be used to identify hosting IPs~\cite{GooglePublicIP,AWSPublicIP,AzurePublicIPList} from Non-hosting IPs. While they may help one identify a subset of hosting IPs, they are not exhaustive due to several reasons: (1) not all cloud services make the IP address ranges readily available to access by the public, and (2) there is a long tail of hosting providers supporting hosting IPs. 
In the prior work, IP address classification is mainly used to efficiently allocate IPs to different organizations and to improve the efficiency of routing IP packets between routers~\cite{Ruiz-Sanchez:2001:IPLookup}. Xie et al.~\cite{Xie:2007:DynamicIP} propose a technique to automatically classify IP addresses as dynamic, that is, DHCP allocated IPs, and static using Hotmail server logs in order to identify spams. Their observation is that most of the spam email servers are hosted at dynamic IP addresses. However, this observation is not reflected on the malicious domains we identified from VirusTotal URL feed and, thus, a different categorization is required. Scott et al.~\cite{Scott:2016:CDNIP} propose to capture the IP footprint of CDN (content delivery network) deployments. While this work identifies IPs of specific shared infrastructures of content delivery, these IPs may not necessarily be public IPs the way we classify in our work in order to study the malicious hosting types. Recently, Nabeel et al.~\cite{inferenceGraph} introduce the concept of shared and dedicated IPs in their domain inference algorithms. We observe a higher number of false positives than reported in their work and we believe this is due to the fact they do not make a distinction between hosting and non-hosting IPs. Recently, Xianghang et al.~\cite{revil:2019}, design a machine learning model to identify residential IP addresses. While their work is useful in identifying potential features, our work is different from theirs as we categorize IPs to identify hosting and non-hosting IPs, where residential IPs are only a subset.

\section{Conclusions}\label{sec:conclusions}
In this paper, we build first machine learning models to identify the hosting types of IP addresses. Specifically, we build a classifier to categorize IP addresses as hosting and non-hosting and a subsequent classifier to categorize hosting IPs as shared or dedicated. From a corpus of 143,316 malicious domains, we observe the concerning trend that attackers 97.1\% of the time utilize shared hosting IPs to host their malicious domains. Further, we show that majority of these shared IPs host many potentially benign domains and only a handful of malicious domains. This makes reputation systems that rely on hosting infrastructure properties less effective. We believe more should be done by hosting providers to safeguard their shared hosting infrastructures.

\end{document}